\documentstyle[twoside,fleqn,espcrc2]{article}

\newcommand{\AmS}{{\protect\the\textfont2
  A\kern-.1667em\lower.5ex\hbox{M}\kern-.125emS}}
\input epsf
\title{
The Role of Monopoles for Color Confinement}
\author{H.~Ichie
\thanks{ichie@rcnp.osaka-u.ac.jp},
A.~Tanaka and H.~Suganuma
\address{Research Center for Nuclear Physics (RCNP), Osaka University\\
Mihogaoka 10-1, Ibaraki, Osaka 567, Japan}%
}

\begin{document}

\begin{abstract}
We study
the role of the monopole for color confinement 
by using the monopole current system.
For the self-energy of the monopole current less than ln$(2d-1)$,
long and complicated monopole world-lines appear
and the Wilson loop obeys the  
area law, and therefore 
 the monopole current system almost
reproduces essential features of confinement properties in the long-distance physics.
In the short-distance physics, however, 
the monopole-current theory would become nonlocal
due to the monopole size effect. This monopole size would provide a 
critical scale of QCD in terms of the dual Higgs mechanism.
\end{abstract}

\maketitle

\section{ Kosterlitz-Thouless Type Transition in Monopole~Current~Dynamics}

Color confinement 
can be interpreted using the dual version of the 
superconductivity\cite{nambu}-\cite{ichiea}.
In this picture, condensation  of color magnetic monopoles  is the key 
concept and leads to squeezing of the color-electric flux between quarks 
through the dual Meissner effect. 
QCD is reduced into an abelian gauge 
theory with monopoles in the 't Hooft abelian 
gauge\cite{thooft}-\cite{suganuma3}. 
As for the maximally abelian (MA) gauge, recent lattice studies suggest that
only abelian gauge fields 
with monopoles are essential for description of nonperturbative 
phenomena in the infrared region\cite{poly}-\cite{suganuma2}.

We study the role of monopoles for confinement using a monopole action.
In the confinement phase, only short-range interactions remain 
among monopoles\cite{suganuma3}
due to the {\it screening effect} \cite{stack} by the
dual Higgs mechanism, and therefore the  infrared monopole-current dynamics
would be described by a simple action\cite{ezawa,ichieb},
$
S$$=$$\alpha^{\rm lat}$$\sum_{s,\mu}$$k_\mu^{\rm lat}(s)^2 
$
in the lattice formalism.
Here,
$k^{\rm lat}_\mu (s)$$\equiv$$e /(4\pi) \cdot a^3 k_\mu(s) \in${\bf Z}  and
$\alpha^{\rm lat}$$\equiv$$\alpha / 2$$\cdot$$(4\pi/ea)^2 $
denote the magnetic current and its self-energy on the lattice, respectively.

The monopole current system is generated with
the current conservation condition, $\partial_\mu k_\mu^{\rm lat} = 0$.
Here, we 
update the {\it link} of the monopole current using the Metropolis method. 

We show in Fig.1 the monopole density 
$\rho_M$$\equiv$$\frac{1}{4V}\sum_{s,\mu} |k^{\rm lat}_\mu(s)|$ 
with the lattice volume $V$$=$$8^4$
and the clustering parameter $\eta 
\equiv \sum_i$$ L^2_i/(\sum_i L_i)^2$,
where $L_i$ denotes the loop length of the $i$-th monopole  cluster.
With increasing $\alpha$,
the monopole density $\rho_M$ is reduced
gradually for $\alpha 
$$\stackrel{<}{\scriptstyle{\sim}}
$$\alpha_c$.
The clustering parameter is drastically changed  at $\alpha_c$.
Quantitatively, the critical value of monopole self-energy,
$\alpha_c$$\simeq$$1.85$, is close to ${\rm ln7}$$\simeq$$1.95$, which is
derived from the entropy factor $(2d-1)^L = 7^L$ in the 4-dimensional 
self-avoiding random walk\cite{ichieb}.
Such a transition is quite similar to the 
Kosterlitz-Thouless type transition in 1+2-dimensional superconductors.
Vortex condensation at  high temperature in the superconductor
corresponds to  monopole condensation in the strong coupling region 
of QCD.

\section{Dual Field Formalism and Role of Monopoles for Confinement}

In this section, we study 
 confinement properties in the monopole current system
with $k_\mu$$\ne$$0$ and $j_\mu$$=$$0$, which is the dual version of QED.
To estimate the abelian Wilson loop, we introduce the dual gauge field
$B_\mu(x)$, since 
the 
gauge field $A_\mu(x)$ 
inevitably includes the singularity as the Dirac string
in the presence of  magnetic monopoles.
Using the relations
$
\partial_{\mu}  {}^*F_{\mu\nu} = k_\nu
$
and
$
{}^*F_{\mu\nu}$$=$$
\partial_\mu B_\nu$$-$$\partial_\nu B_\mu,
$
the dual gauge field $B_\mu$ can be described from the monopole current $k_\mu$,
\begin{equation}
B_\mu (x) = \partial^{-2} k_\mu(x) = -\frac{1}{4\pi^2}
\int d^4 y \frac{k_\mu(y)}{|x-y|^2}, 
\label{eq:dual}
\end{equation}
in the dual Landau gauge $\partial_\mu B_\mu = 0$.
The Wilson loop is obtained as
$
\langle W \rangle 
 =  \langle {\rm exp} 
\{ {ie \frac12 \epsilon_{\mu\nu\alpha\beta} 
 \int d\sigma_{\mu\nu}
(\partial_\alpha B_\beta - \partial_\beta B_\alpha) 
} \} \rangle.
$
In estimating the integral in Eq.(\ref{eq:dual}) numerically, we  should take 
{\it a finer mesh than original lattice spacing $a$} 
to extract $B_\mu$ correctly.
Otherwise, the unexpected electric currents appear as the lattice artifact.
However, too finer mesh is not necessary, because 
the original current $k^{\rm lat}_\mu(s)$ includes the error 
in the order of $a$.
Numerical analyses show that
the calculation with $c \le a/2$ is good enough for the 
estimation of $B_\mu$, and hence we take $c=a/2$ hereafter.
On lattices, the dual gauge field 
$\theta_\mu^{\rm dual} \equiv a e B_\mu /2$ 
is defined in the dual Landau gauge;
$
\theta_\mu^{\rm dual} (s+\mu) \equiv 
2 \pi \sum_{s'}\Delta(s-s') k^{\rm lat}_\mu(s') 
$
using the lattice Coulomb propagator $\Delta(s)$.
The dual version of the abelian field 
strength $\theta_{\mu\nu}^{\rm dual} \equiv e a^2 {}^*F_{\mu\nu}/2$ 
is defined by
$
\theta_{\mu\nu}^{\rm dual}(s) \equiv
\partial'_\mu \theta^{\rm dual}_\nu(s) 
- \partial'_\nu \theta^{\rm dual}_\mu(s),
$
where $\partial'$ denotes the backward derivative.

%
%

As shown in Fig.2, the Wilson loop  $\langle W \rangle$ exhibits the
area law, which leads to the linear confinement potential.
In Fig.3, we show the string tension 
$\sigma^{\rm lat} \equiv \sigma^{\rm phys} a^2$ in the lattice unit.
For
$\alpha$$ 
\stackrel{<}{\scriptstyle{\sim}}
$$\alpha_c$,
the string tension is finite due to the long monopole currents,
the current simulations for 
$\alpha$$ 
\stackrel{>}{\scriptstyle{\sim}}
$$\alpha_c$ indicate
that $\sigma^{\rm lat}$$\simeq$$0$  
in the system without monopole currents.
On the other hand, lattice QCD simulations show that the long monopole 
currents cover the whole system and string tension is finite in the 
confinement phase even for the large $\beta$ as $\beta = 2.3$$\sim$$2.5$
on $16^4$ lattices.
Therefore, the monopole self-energy $\alpha$ extracted from QCD must be 
smaller than $\alpha_c$$\equiv$ ln($2d-1$) = ln7 in the confinement phase.
Here, to extract the monopole action, one should use the monopole part in
QCD, where only monopole currents exist.
Since the electric current and the short-range Coulomb force cannot be 
expressed only by the monopole currents, it is dangerous to compare the 
monopole action with the full SU(2) QCD (or the abelian part);
the former includes monopole currents only, while the latter includes also
electric currents and Coulomb force.
 In any case, {\it  the monopole action is to be extracted from the monopole part
in QCD and the monopole self-energy $\alpha$ should be smaller than 
$\alpha_c$=ln7 in the confinement phase. }

\section{Monopole Size Effect and Critical Scale in QCD}

The aim of our study is to analyze the theoretical structure of
QCD in terms of the monopole degrees of freedom.
Our studies indicate that the multi-monopole system has a similar
structure to QCD in the MA gauge.
However, the monopole in abelian-projected QCD
has the underlying structure reflecting QCD.
For instance,
QCD-monopoles would be as  {\it extended objects}
like hadrons\cite{suganuma3}, 
because it is not an elementary
particle in QCD, but a  {\it collective mode composed by gluons}.
We compare the monopole current system with 
QCD \cite{poly} especially in terms of the self-energy 
and the size of monopoles\cite{ichie}. 

In the static frame of the monopole
with an intrinsic radius $R$, 
the spherical magnetic field is created around the monopole as 
%
%
$
{\bf H}(r)  =
  {g(r) \over 4\pi r^3} {\bf r}={{\bf r} \over e(r) r^3} 
$
for $ r \ge R $
and
$
{\bf H}(r)  =  
{{\bf r} \over e(r) R^3} 
$
for $r$$\le$$R$,
with QCD running gauge coupling $e(r)$\cite{ichieb}.

For $a$$>$$R$, 
the electromagnetic energy observed on the lattice around 
a monopole is roughly estimated as 
$
M(a)$$\simeq$$\int_a^\infty d^3x {1 \over 2}{\bf H}(r)^2 
$$\simeq$$ 
{2\pi \over e^2(a)a},
\label{eq:massa}
$
which  largely depends on the lattice mesh $a$.
Hence, the monopole contribution to the 
lattice action reads $S$$=$$M(a)a$$\cdot$$L$, 
where  $L$  denotes  the monopole current length  
measured in the lattice unit $a$. 
Therefore, $M(a)$ is  related to the monopole-current coupling 
$\alpha^{\rm lat}$ 
 and $\beta$$=$$2N_c/e^2$, 
$
\alpha^{\rm lat}$$\simeq$$M(a)a$$\simeq$${2\pi \over e^2(a)}
$$=$${\pi \over 2}$$\cdot$$\beta_{\rm SU(2)}.
$
By the screening effect, this relation becomes 
$
\alpha^{\rm lat}$$\simeq$$ 
{\pi \over 4}$$\cdot$$\beta_{\rm SU(2)}
$\cite{ichieb},
which is consistent with the numerical result
as shown in Fig.3.
Thus, as long as the mesh is large as $a$$>$$R$, the lattice monopole 
action 
would not need modification by the monopole size effect, 
and  $\alpha^{\rm lat}$ is proportional to $\beta$.

On the other hand, 
there exists a critical coupling $\alpha_c$$\simeq$${\rm ln}$7
in the current dynamics. 
Above $\alpha_c$, the lattice current action 
provides 
no monopole condensation and no confinement, while 
$\beta$$\rightarrow$$\infty$ can be taken in the original QCD 
keeping the confinement property shown in Fig.3. 
Such a discrepancy between $\beta$ and $\alpha$ 
can be naturally interpreted by introducing the monopole size effect,
which should lead the nonlinear correspondence between $\alpha$
and $\beta$.
For $\alpha$$<$$R$, the extended structure of the monopole can be observed 
on the lattice,
and the monopole creates the electromagnetic energy 
$
M(a)$ 
$\simeq$${g^2(R)\over 8\pi R}$$=$${2\pi \over e^2(R) R}
$
on the lattice, which is almost $a$-independent. 
Accordingly, the lattice action should be changed as 
$
S$$=$$M(R) a \sum_{(s,s'),\mu} k_\mu(s)k_\mu(s')$$\cdot$$
\theta \left( {R \over a}-|s-s'| \right) \delta_{s_\mu, s'_\mu} 
$
because of the self-avoidness originating from the monopole size $R$.
Thus, the monopole-current theory becomes nonlocal 
in the UV region.

The monopole size $R$ provides a critical scale 
for the description of QCD in terms of the dual Higgs mechanism. 
In the infrared region as $a$$>$$R$, 
QCD can be approximated as a local monopole-current action\cite{ezawa}, 
and the QCD vacuum can be regarded as the dual superconductor. 
On the other hand, in the ultraviolet region as $a$<$R$, 
the monopole theory becomes nonlocal and complicated 
due to the monopole size effect, and the perturbative QCD 
would be applicable instead. 

We acknowledge Prof. H.Toki and Dr. S.Sasaki for useful discussions.
H.I. is supported by 
Research Fellowships of the 
Japan Society for the Promotion of Science 
for Young Scientists.

\hspace{2cm}

\noindent {\bf \large Figure Captions}

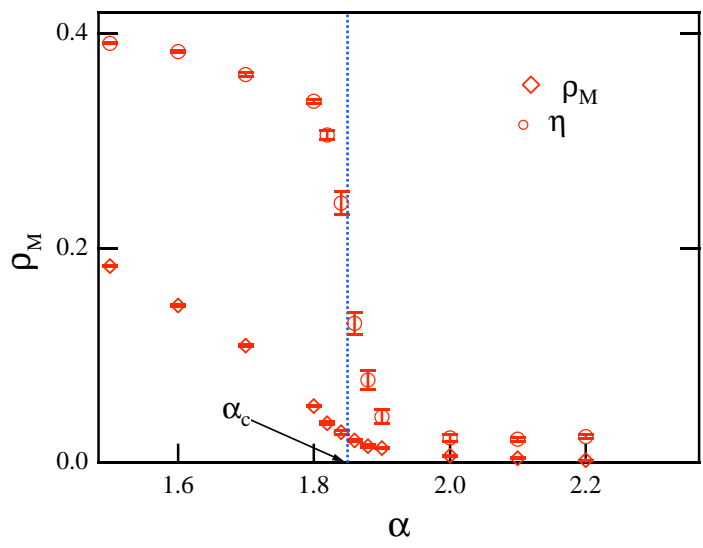
\noindent {\bf Fig.1}
The monopole density $\rho_M$  and
the clustering parameter $\eta$ as the function of $\alpha$.

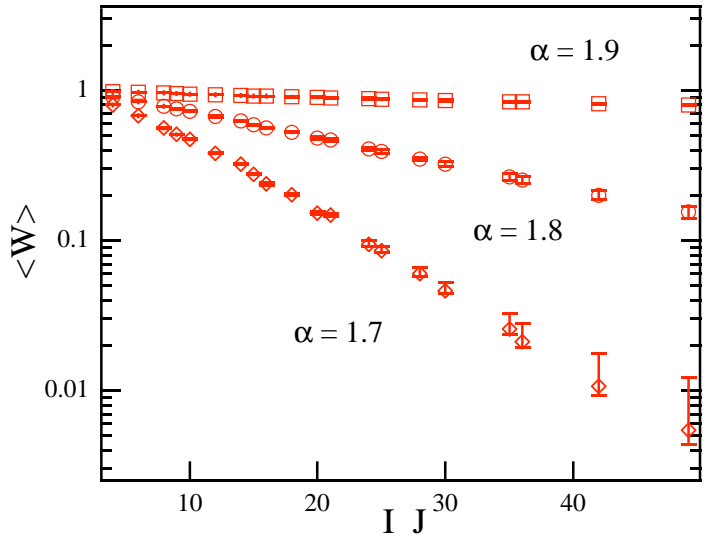
\noindent {\bf Fig.2}
The Wilson loop 
$\langle W(I \times J) \rangle$ on the lattice with the mesh $c=a/2$
for $\alpha=1.7,1.8,1.9$.

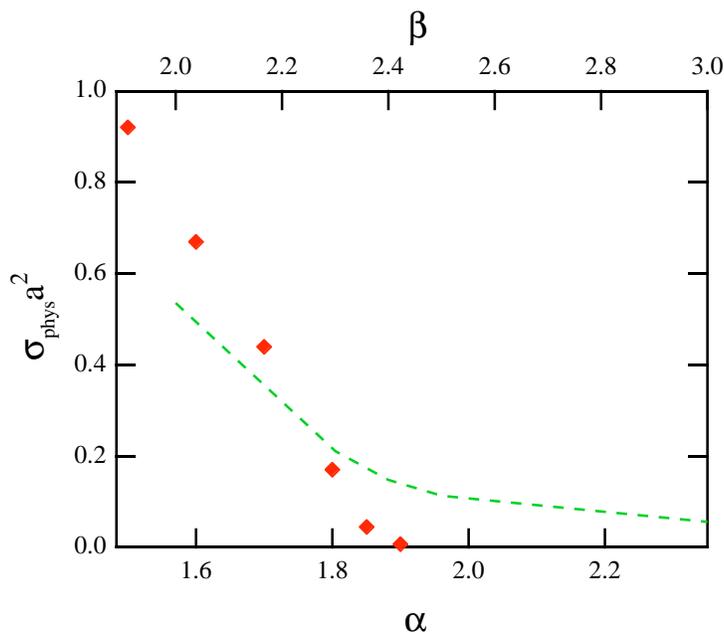
\noindent {\bf Fig.3}
 The string tension $\sigma^{\rm lat} \equiv \sigma^{\rm phys} a^2$
as the function of $\alpha$ in the multi-monopole system.
The dotted line denotes the Creutz ratio 
in the lattice QCD with $\beta = \frac{4}{\pi} \alpha$.

\end{document}